\documentclass[12pt,preprint]{aastex}
\usepackage{emulateapj5}

\makeatletter

\newenvironment{inlinefigure}{%
\def\@captype{figure}%
\noindent\begin{minipage}{0.999\linewidth}\begin{center}}
{\end{center}\end{minipage}\smallskip}
\makeatother

\def\asca       {{\em ASCA}\/}
\def\chandra    {{\em Chandra}\/}
\def\ginga      {{\em Ginga}\/}
\def\einstein   {{\em Einstein}\/}
\def\xmm        {XMM-{\em Newton}\/}
\def\rosat      {{\em ROSAT}\/}

\def\mydegree{$^\circ\mskip-5mu$}
\def\myarcmin{$^\prime\mskip-5mu$ }
\def\myarcsec{$^{\prime\prime}\mskip-5mu$}

\usepackage{mathptmx}

\begin{document}

\title{\emph{Chandra} observations of the galaxy cluster A478: the interaction
of hot gas and radio plasma in the core, and an improved determination
of the Compton y-parameter}

\author{
M.\ Sun,$^{\!}$\altaffilmark{1}
C.\ Jones,$^{\!}$\altaffilmark{1}
S. S. Murray,$^{\!}$\altaffilmark{1}
S. W. Allen,$^{\!}$\altaffilmark{2}
A. C. Fabian,$^{\!}$\altaffilmark{2}
A. C. Edge,$^{\!}$\altaffilmark{3}
} 
\smallskip

\affil{\scriptsize 1) Harvard-Smithsonian Center for Astrophysics,
60 Garden St., Cambridge, MA 02138; msun@cfa.harvard.edu}
\affil{\scriptsize 2) Institute of Astronomy, Madingley Road, Cambridge
CB3 0HA, UK}
\affil{\scriptsize 3) Department of Physics, University of Durham,
South Road, Durham DH1 3LE, UK}

\begin{abstract}

We present the results from a 42 ks \chandra\ ACIS observation
of the galaxy cluster A478. This cluster is generally considered
to be highly relaxed. The \chandra\ image reveals, for the
first time, X-ray cavities in the hot gas within the central 15
kpc radius of A478. Two weak and small ($\sim$ 4 kpc) radio lobes
that extend from the central nucleus, are detected in a 1.4 GHz
VLA observation. The radio lobes are roughly along the direction
of the X-ray cavities, but are much smaller than the X-ray cavities.
We propose a ``donut'' configuration for the hot gas within the
central 15 kpc, created by the interaction of the gas with the
radio plasma that originated from the nucleus. The current radio
activity of the central radio source is weak ($\sim$ 0.2\% of
Hydra A) and the total radio power is at least 10 times smaller
than the minimum power needed to create the cavities. We compare
A478 with other galaxy clusters where similar X-ray cavities were
found. A478 and A4059 host much weaker central radio sources than
do others with similar size X-ray cavities. On larger scales,
deprojected temperature and density profiles are obtained for
A478. We used these to derive the Compton y-parameter, for the
first time, through direct integration. The result has a much
smaller statistical uncertainty than previous ones. This serves
as an example of how high quality X-ray data can help constrain
H$_{0}$. The corresponding H$_{0}$ also was derived combining the
available Sunyaev-Zeldovich effect (SZE) measurement.

\end{abstract}


\section{Introduction}

A478 is an X-ray-luminous cluster at a redshift of 0.088 (Zabludoff,
Huchra \& Geller 1990). Previous X-ray observations with \ginga, \einstein,
and \rosat\ suggested that it contains one of the largest central cooling
flows with estimates up to
$\sim$ 1000 M$_{\odot}$ yr$^{-1}$ (Johnstone et al. 1992; Allen et al. 1993;
White et al. 1994; Allen 2000). These observations also indicated excess X-ray
absorption above the Galactic value.
The X-ray morphology of A478 implies it is a very relaxed cluster (e.g.,
Allen et al. 1993), although it is elongated in the Northeast-Southwest
direction (e.g., White et al. 1994). This elongation is along the same
direction as the cluster galaxies (Bahcall \& Sargent 1977), as well as the light
distribution of the cD galaxy (White et al. 1994). A478 also has a marginal CO
detection of 4.5$\pm$2.2 $\times$ 10$^{9}$ M$_{\odot}$ (Edge 2001). A strong SZE decrement
is detected in this cluster, making it an excellent target to measure
the Hubble constant (Myers et al. 1997, M97 hereafter; Mason et al. 2001,
M01 hereafter).

In this Letter, we present initial results from a recent \chandra\ 42 ks
ACIS observation of A478, with emphasis on the central X-ray gas/radio
source interaction and an improved result for the predicted Compton
y-parameter using the
temperature and density profiles measured with \chandra. A detailed
analysis of the cooling flow will be presented in another paper. 
Throughout this paper we assume H$_{0}$ = 70 km s$^{-1}$ Mpc$^{-1}$,
$\Omega$$_{\rm M}$=0.3, and $\Omega_{\rm \Lambda}$=0.7 unless specified.
At a redshift of
0.088, the luminosity distance to A478 is 403 Mpc and 1$''$=1.65 kpc.

\section{\chandra\ observation \& data reduction}

A478 was observed for 42.4 ksec on January 29, 2001 by \chandra\ with
the Advanced CCD Imaging Spectrometer (ACIS). The observation used
CCDs I2-I3 and S1-S4 with the cluster centered on S3.
The data were telemetered in Faint mode and \asca\ grades 1, 5 and 7
were excluded. Known bad columns, hot pixels, and CCD node boundaries
also were excluded. We investigated the background light curves
from the back-illuminated (BI) S1 CCD and from the
front-illuminated (FI) CCDs, using the standard software$^{\rm 1}$. No
background flares were found in this observation. The streak events in the
S4 CCD were removed by CIAO DESTREAK software.

In our analysis we used period D blank field background data by M. Markevitch
\footnote{http://cxc.harvard.edu/contrib/maxim/bg/index.html}.
The particle background level was measured in the hard
band (7 - 11 keV) for the S1 CCD (far from the cluster) and in PHA channels
2500-3000 ADU for all CCDs. We found that the particle background level
was 10\% lower than that of the period D background data. This is within the
uncertainties for the background files. Moreover, the observation was taken
at the time that S3 background rate was the lowest
\footnote{http://cxc.harvard.edu/ccw/proceedings/presentations/m\_markevitch/pg002.html}.
Thus, we decreased the background normalization by 10\%
to fit the particle background level at the time of the A478 observation
as recommended in Markevitch et al. (2002).
This correction has only a small effect on the best-fit temperature (e.g.,
10\% for the outermost annulus in $\S$4).
Note that although the correction should apply only to the particle
component of the total background and not to the cosmic x-ray background or
CXB, the particle component is dominant at energies greater than $\sim$ 3 keV
which are most important for measuring the gas temperature. This small
rescaling of the CXB component has little effect on the spectral
fitting (e.g., $\sim$ 5\% for the outermost annulus). We also corrected
for the readout effect in the S3 CCD caused by the bright cluster
core (see Markevitch et al. 2000 for details). Two corrections were made
to the ACIS low energy quantum efficiency (QE). The first uses CXC (\chandra\
X-ray Center) contributed software to correct for the QE degradation
\footnote{http://cxc.harvard.edu/cal/Links/Acis/acis/Cal\_prods/qeDeg/index.html} (Plucinsky et al. 2002).
The second corrects the QE by an empirical factor of 0.93 below 1.8 keV
in the FI CCDs to improve the cross-calibration with the BI CCDs
\footnote{http://asc.harvard.edu/cal/Links/Acis/acis/Cal\_prods/qe/12\_01\_00/}
(also see Markevitch \& Vikhlinin 2001 for details). The tools calcrmf
and calcarf by Vikhlinin were used to generate response files.

The calibration files used correspond to CALDB 2.15 from the CXC. The
uncertainties quoted in this paper are 90\% confidence intervals
unless specified. The solar abundance table by
Anders \& Grevesse (1989) and a cluster redshift of 0.088 were
used in the spectral fits and computations.

\section{Central region of A478}

On large scales from 50 kpc to several Mpc, A478 appears quite relaxed,
although the X-ray morphology is elliptical rather than
circular (Fig. 1a). The superior angular resolution of \chandra\ allows
us to examine structures on scales of several kpc at the center of A478.
In the central 20 kpc (Fig. 1b), the cluster shows significant structures.
The X-ray emission appears double-peaked (separated by $\sim$ 8 kpc) with
the two peaks lying east and west of the cD galaxy nucleus.
In contrast to the double peaks, there are depressions North and
South of the cD nucleus. To better quantitatively show the depressions,
we measured the surface brightness profiles in four directions from the
nucleus within $\sim$ 15 kpc (Fig. 2). The surface brightness in the North
and South sectors is 20\% - 30\% lower than that to the West and East.
Beyond $\sim$ 20 kpc, no significant difference between the surface
brightness of the sectors is seen.
Since the X-ray depressions seen in A478 are similar
to the X-ray cavities observed in other clusters (e.g., Hydra A, McNamara
et al. 2000) where central radio sources are certainly responsible, we
investigated the radio activity of the central source in A478.

A 1.4 GHz VLA image of A478, obtained in 630 s with the A array on March
1, 1998, is shown as isointensity contours in Fig. 1b. The central radio
source is composed of a bright point-like source at the position of the
cD nucleus, with weak extensions (lobes) towards the Northeast
and Southwest. The measured 1.4 GHz total flux is about 30 mJy, while the
lobes contribute $\sim$ 35\% of this. The radio lobes are quite
small: $\sim$ 4 - 5 kpc extension for the Southwest one and $\sim$ 3 - 4
kpc extension for the Northeast one (the VLA angular resolution at 1.4 GHz
is 1.5$''$, or 2.5 kpc). They lie roughly along the direction of the X-ray
depressions, although the X-ray depressions are much larger than the
detected radio lobes. The two X-ray peaks are located along the
East and West sides of the radio source. The correspondence of radio
lobes and X-ray structure suggests that the X-ray structure was created
by the interaction of the hot gas with the radio lobes. Similar X-ray
structures (X-ray cavities) have
been found at the centers of a number of cooling flow clusters (Hydra A -
McNamara et al. 2000; Perseus cluster - Boehringer et al. 1993, Fabian et
al. 2000; A2052 - Blanton et al. 2001; A2597 - McNamara et al. 2001; A4059
- Heinz et al. 2002). Generally, they are thought to form when the hot
gas is displaced by the expansion of radio lobes or rising buoyant bubbles
(e.g., Churazov et al. 2001; Soker, Blanton \& Sarazin 2002). For
consistency, hereafter we also call the X-ray structure in A478 cavities.

No X-ray point source is detected at the position of the nucleus
of the cD galaxy, where a radio point source is detected (Fig. 1b).
Assuming a $\Gamma$=1.7 power law modified by an absorption of
2.6$\times$10$^{21}$ cm$^{-2}$ (the best-fit, see $\S$4) and a cluster
redshift of 0.088, the 3 $\sigma$ upper limit on the point source luminosity
is 3$\times$10$^{41}$ ergs s$^{-1}$ (0.5 -
10 keV). If the central source is as highly obscured as the one in Hydra A
(McNamara et al. 2000), the upper limit will be higher (e.g.,
1.1$\times$10$^{42}$ ergs s$^{-1}$ for N$_{\rm H}$=2.6$\times$10$^{22}$
cm$^{-2}$, which is comparable to the source detected in Hydra A).

Spectra of the two X-ray peaks were extracted from 10$''$ (major axis) $\times
4''$ (minor axis) ellipses centered on them. We fit each with a
MEKAL model and found consistent temperatures and abundances: 3.2$\pm$0.2
keV and 0.60$^{+0.20}_{-0.17}$ for the eastern peak, 3.1$\pm$0.2 keV and
0.65$^{+0.21}_{-0.18}$ for the western peak. A hardness ratio map (2 - 5 keV
/ 0.5 - 2 keV) for the central 60 kpc$\times$60 kpc region is shown in
Fig. 3. The coolest region contains the two peaks and regions between them.
The overall temperature increases outwards from the core. There is no
indication at shock heated regions around the X-ray cavities.
The spectrum of the central 20 kpc (radius) region was fit with a MEKAL +
MKCFLOW model to search for any cooling flow component.
The inclusion of a cooling flow model (MKCFLOW) did not significantly
improve the fit, if we allow the gas to cool to zero. However, if we
use a lower temperature limit of $\sim$ 1 keV, the fit improves
significantly. Thus, while there is a clear signature of gas cooling
towards the center (see $\S$4), the spectral fits do not require gas at
very low temperatures.

\section{Analysis of the temperature, absorption and heavy element abundance profiles}

We first compare the integrated spectrum of A478 from the entire S2, S3 and
S4 CCDs with previous results. Data in the 0.7 - 8 keV energy band
are used. The results are N$_{\rm H}$ = (0.259$\pm$0.003)$\times$10$^{22}$
cm$^{-2}$, T = 7.18$\pm$0.11 keV, and abundance 0.37$\pm$0.02 solar.
From \asca\ GIS 1 - 9 keV data, White (2000) obtained a temperature
of 6.6$\pm$0.4 keV and an abundance of 0.31$\pm$0.05
with a fixed absorption of 3$\times$10$^{21}$ cm$^{-2}$. If we fix the
absorption at their value, we find a consistent temperature T = 6.62$\pm$0.09
keV and an abundance of 0.37$\pm$0.02.

Although in X-rays A478 is somewhat elliptical, it is useful to derive radial
properties of the temperature, absorption and heavy element abundance.
For the profiles, we centered all annuli at
R.A. = 04$^{h}$13$^{m}$25$^{s}$.3, decl. = 10\mydegree~27\myarcmin53\myarcsec
and required each annulus to contain 6000 - 8000 source counts. Each
annulus was fit by a MEKAL model, with the absorption, temperature
and abundance as free parameters. The reduced $\chi^{2}$ ranged
from 0.8 to 1.4 and most were around 1.0 (for 125 - 245 degrees of freedom),
which is acceptable considering the large number of source counts and the
current ACIS calibration. The derived temperature profile is shown in Fig. 4.
The temperature falls from $\sim$ 8.5 keV at large radii to $\sim$ 3 keV
at the center. Between radii of $\sim$ 110 kpc and $\sim$ 500 kpc, gas
temperatures are consistent with a constant temperature of $\sim$ 8.5 keV.
To better constrain the absorption and abundance, we used larger radial bins
and obtained the results shown in Fig. 5. The abundance increases from $\sim$ 0.3
solar at several hundred kpc to $\sim$ 0.6 solar at the center.
The absorption also increases towards the center. The spectra in the regions
outside $\sim$ 500 kpc (in CCDs S2, S4, I2, and I3) were fitted
simultaneously with a fixed absorption of 0.24$\times$10$^{22}$ cm$^{-2}$
(extrapolated from the absorption profile shown in Fig. 5).
The best-fit temperature is 6.45$^{+0.80}_{-0.75}$ keV and will be taken
as the ``background'' temperature in the deprojection analysis ($\S$5).

\section{Deprojection analysis}

Although the X-ray morphology of A478 is elliptical rather than circular,
the ellipticity is small ($\sim$ 0.2) and a spherical approximation
serves as a simple way to do the deprojection and other analyses.
Thus, based on the 2D temperature profile obtained in $\S$4, we performed
a deprojection analysis to obtain the 3D temperature profile and abundance
profile (see e.g., David et al. 2001 for details of the spectral deprojection).
Nine radial bins were chosen to do the spectral deprojection.
These were the same ones used above to derive the absorption and abundance
profiles.
To account simply for the slight decrease of absorption with radius, we
used two absorption values, 0.28$\times$10$^{22}$ cm$^{-2}$ for annuli 1-3
and 0.25$\times$10$^{22}$ cm$^{-2}$ for annuli 4-9 (Fig. 5), derived by
fitting a single absorption for these regions ($\chi^{2}$ = 8.1/7).
The deprojection result is shown in Fig. 4. For regions within a
radius of $\sim$ 110 kpc, we also fit the 3D temperatures with a power law
given by T(r) = T$_{0}$(r/10kpc)$^{\alpha}$. The best fit is T$_{0}$ =
2.79$\pm$0.09 keV and $\alpha$ = 0.412$\pm$0.017 (1$\sigma$ errors).
Deprojected abundances are shown in Fig. 5.

The surface brightness profile of A478 is shown in Fig. 4. There
is at least one inflection point at $\sim$ 40 kpc. If the surface brightness
profile is fit by a single $\beta$-model, the central excess is quite
significant and the fit is poor. Thus, we use a deprojection technique
to determine the electron density profile rather than fitting with a
multivariate function (see a recent example in David et al. 2001).
This technique converts the observed surface brightness to the electron
density starting in the largest annulus and then
determines the density at progressively smaller radii, after removing the
projected emission from larger radii. The fraction of cluster emission
in all radial bins arising from gas outside the outermost annulus ($\sim$
5$'$ radius) is corrected using the power law fit to both \chandra\ and
PSPC surface brightness profiles in regions outside the outermost annulus.
The deprojected temperature and abundance profiles were combined to produce
an emissivity profile (with uncertainties) used in the conversion.
We performed 400 Monte Carlo simulations to estimate the errors.

The deprojected electron density profile is shown in Fig. 4, as well
as the pressure profile. The density
profile gradually steepens from n$_{\rm e} \propto$ r$^{-0.25}$ at 10 kpc
to n$_{\rm e} \propto$ r$^{-1.65}$ at 250 kpc, and there is no flat density
core. The isobaric cooling time is 4$\times$10$^{8}$ yr in the innermost
10 kpc and less than a Hubble time ($\sim$ 10$^{10}$ yr) within the
central 140 kpc. The pressure reaches a maximum at r $\sim$ 15 kpc, at
roughly the outer edge of the X-ray cavities.

\section{Discussion}

\subsection{Central X-ray gas / radio source interaction}

As shown in $\S$3, there are two X-ray cavities in A478. The alignment of
the radio lobes and X-ray cavities implies that the X-ray gas was swept
out of the cavities by expanding radio plasma. We suggest that the X-ray
double peaks result from a ``donut'' configuration for the hot gas around the
central nucleus, with the plane of the ``donut'' perpendicular to the
direction of the radio lobes, while the central parts of the ``donut''
were displaced by the expanding radio plasma. The absence of shocked
regions in the A478 core (Fig. 3) is consistent with the radio plasma
expanding subsonically, which is similar to what was found in other clusters (e.g., Hydra A
and Perseus). If we assume the X-ray cavities are shaped like circular cones,
the minimum energy required to sweep the gas out of the current
cavities is pV $\backsimeq$ 3 $\times$ 10$^{58}$ ergs. Churazov et al.
(2001) showed that a large and strongly underdense bubble will rise
with a velocity comparable to the Keplerian velocity. Therefore, as a first order
approximation, we take the Keplerian velocity at 10 kpc to represent the
rising velocity of the bubble ($\sim$ 460 km/s, the mass is estimated from
the deprojected temperature and density under the assumption of hydrostatic
equilibrium). Thus, the time to form the current cavities is $\sim$ 3
$\times$ 10$^{7}$ yr. Therefore, the minimum average power from the central source
to create the X-ray cavities is $\sim$ 3 $\times$ 10$^{43}$ ergs s$^{-1}$,
which is at least 10 times larger than the current total radio power from
the central source. Since the total radio power is simply a lower limit
to the kinetic power of the jet (e.g., Perseus --- Fabian et al. 2002),
more radio information is required to conclude whether the current
kinetic power of the jet could
create the X-ray cavities. However, the observed radio lobes
are considerably smaller than the X-ray cavities, which may imply a
fading of the radio source within $\sim$ 10$^{7}$ --- 10$^{8}$ yr.

Five other clusters also contain X-ray cavities.
Although each is very complicated, in Fig. 6 we compare the cavity
size and the 1.4 GHz luminosity of the central radio source for all of
them. Although the kinetic power of the radio source is generally unknown,
Heinz et al. (2002) argued that the upper limit on the current kinetic
power was proportional to the current 1.4 GHz luminosity. In A478 the
1.4 GHz luminosity of the central source is the smallest in this sample
(less than 7\% of any others) and the radio lobes are the smallest.
Thus, A478 serves as an important example for understanding
X-ray gas / radio plasma interaction in the center of cooling-flow
clusters. The size of the X-ray cavities in A478 are $\sim$ 3/4 the size
of those in Perseus (Fig. 6), but their 1.4 GHz luminosities differ
by $\sim$ 20. In Fig. 6, if we use the three clusters where the central radio
sources are powerful enough to create the X-ray cavities (Hydra A, A2052
and Perseus) to define a line, A478 and A4059 have much lower 1.4 GHz
luminosities ($\sim$ 10\%) than the values predicted by that line.
Heinz et al. (2002)
suggested that the central radio source in A4059 had faded by about
an order of magnitude on a timescale less than 10$^{8}$ yr, since the
current kinetic power they estimated is too small to create the X-ray
cavities. If we adopt their method, A478 is very similar to A4059: in both
clusters, the estimated current kinetic power is too small to create the X-ray
cavities. Thus, A478 and A4059 may be in a different ``phase'' of their
active nuclei cycles than the other sources.

\subsection{An improved estimation of the Compton y-parameter}

To date, X-ray observations have been combined with SZE measurements to estimate
H$_{0}$ for over 20 galaxy clusters (e.g., M97;
Hughes \& Birkinshaw 1998; Reese et al. 2000; Mauskopf et al. 2000).
However, these analyses generally used only single temperatures (or simple
temperature profiles) and density
profiles derived by simple $\beta$-models. These can introduce significant
errors in some cases, such as for cooling flow clusters where a single
$\beta$-model usually fails to describe the central region,
or clusters with significant temperature gradients (Hughes \& Birkinshaw
1998). It is ideal and more accurate to apply directly the observed density
and temperature profiles to models. In this paper, for the first time,
we apply the deprojected density and temperature profiles and use direct
integration to derive the Compton y-parameter
\begin{equation}  \label{y}
y = \frac{k\sigma_T}{m_e c^2} \int\limits_{-\infty}^{\infty} T_e n_e
  \, d\zeta \,.
\end{equation}

In this section, we use H$_{0}$ = 100 h km s$^{-1}$ Mpc$^{-1}$ and
assume spherical symmetry of the cluster X-ray emission. In $\S$5, we
derived the deprojected density profile within the inner 5$'$ (Fig. 4).
For the outer regions, we combine \chandra\ and \rosat\ PSPC images and fit
the surface brightness profile with a $\beta$-model. The derived parameters
are $\beta$ = 0.68$\pm$0.01 and core radius r$_{0}$ = 1.02$\pm$0.07$'$ (1 $\sigma$ errors).
The density profile beyond 5$'$ was obtained from this $\beta$-model.
Combining the deprojected density profile with the deprojected temperature
profiles (n$_{e}$(r) and T$_{e}$(r) in $\S$5), the Compton y-parameters can
be computed for different lines of sight. The uncertainties are estimated from
Monte Carlo simulations. M97 and M01 measured the Compton
y-parameter based on their Owens Valley 5.5 m data at 32 GHz. To compare the
predicted y-parameter from our X-ray observations with their measured value,
we correct for specifics of the 5.5 m telescope (the OVRO beam and the
switching pattern; see eq. 32 - 37 of M97 for details). After those
corrections, we obtain a predicted y-parameter specific for the 5.5 m
telescope: (6.16$\pm$0.06) $\times$ 10$^{-5}$ (y$_{\rm pred}$, 1 $\sigma$
error). The systematic errors from X-ray observations can further
contribute up to 4\% error (mainly from the uncertainties of temperatures).
This value has a much smaller statistical uncertainty than the estimates
in M97 and M01 (4.2$\pm$1.0 $\times$ 10$^{-5}$ and 6.05$\pm$0.54 $\times$
10$^{-5}$). Moreover, our derivation of the Compton y-parameter is
more accurate than the simple methods adopted in M97 and M01,
which were limited by the X-ray data at that time.

We can use this result to obtain an estimate of H$_{0}$.
M01 obtained the observed y-parameter: (7.77$\pm$0.58) $\times$ 10$^{-5}$
(1 $\sigma$ error). However, after we examined the relativistic correction
they used, which depends on the measured temperature, we changed the
relativistic correction factor to 1.026 following Itoh, Kohyama \& Nozawa
(1998). Thus, the observed y-parameter should be (7.71$\pm$0.57) $\times$
10$^{-5}$ (y$_{\rm obs}$, 1 $\sigma$ error). In principle, h =
(y$_{\rm pred}$ / y$_{\rm obs}$)$^{2}$. Including the uncertainty
from intrinsic CMB anisotropies (see details in M01), we obtain a Hubble
constant H = 64$^{+32}_{-18}$ km s$^{-1}$ Mpc$^{-1}$ (1 $\sigma$ error).
Intrinsic CMB anisotropies and the uncertainty of the radio observation
are the main sources of the statistical uncertainties. Besides X-ray
calibration uncertainties mentioned above, other systematic
uncertainties include calibration uncertainties in the radio observations
(3\% - M01), uncertainty from peculiar motion (2\% for 500 km s$^{-1}$
peculiar motion), clumping of X-ray gas ($\sim$ -15\%), undetected radio
sources ($\sim$ 10\%), and unknown geometry ($\sim$ 10\% based on the
analysis in Hughes \& Birkinshaw, 1998). A478 only serves as an example.
As more SZE clusters are observed by \chandra\ and \xmm, and as better
radio observation are available, a systematic analysis will provide a
better constraint on H$_{0}$.

\acknowledgments

The results presented here are made possible by the successful effort of the
entire \emph{Chandra} team to build, launch, and operate the observatory. We
acknowledge helpful discussions with L. David, D. Harris, M. Markevitch, and
A. Vikhlinin. We thank the Smithsonian Institution for support. We thank the
referee for comments. This study was also supported by NASA contract
NAS8-38248.

\begin{figure*}
\vspace{-8cm}
  \centerline{\includegraphics[height=1.5\linewidth]{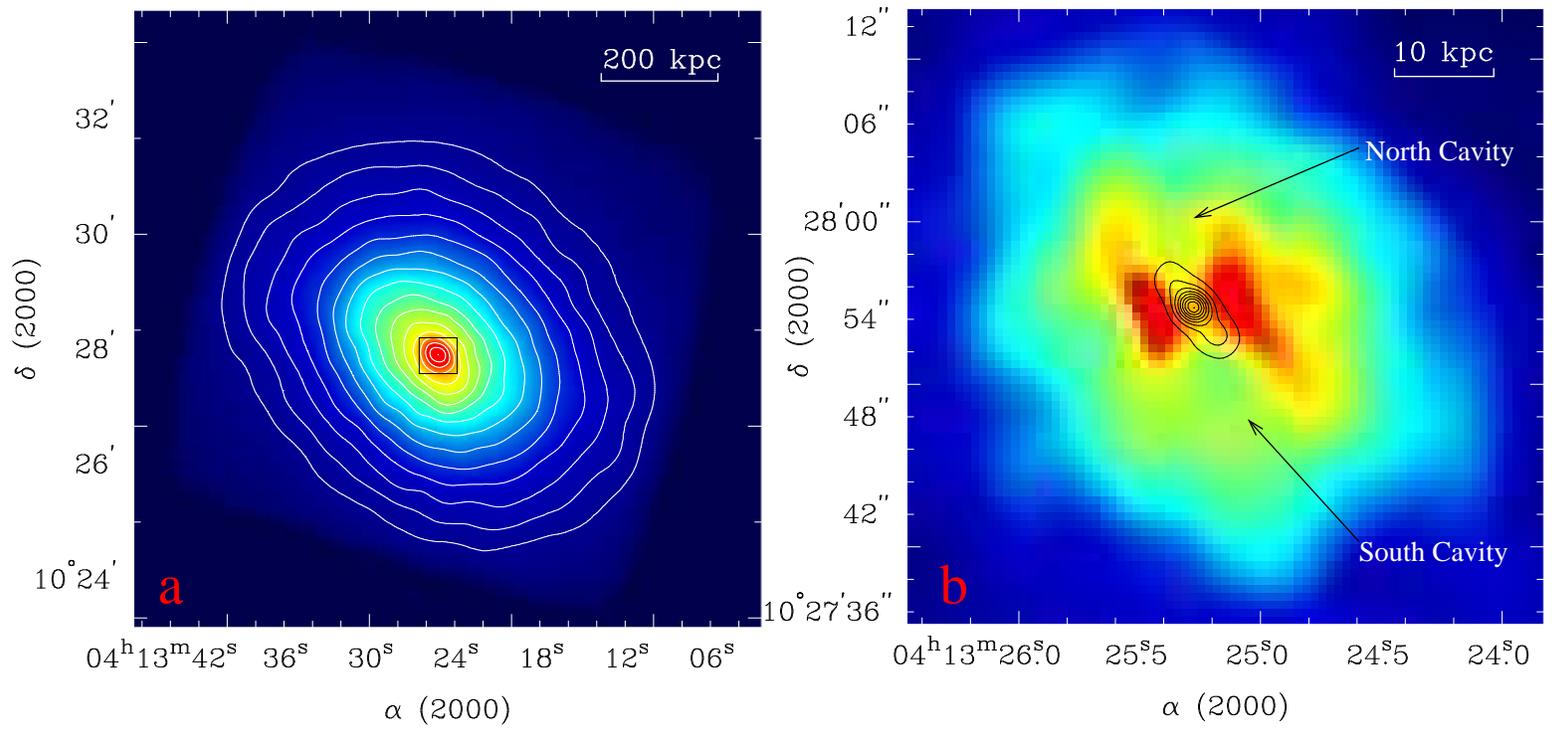}}
\vspace{-8.5cm}
  \caption{{\bf (a)}: \chandra\ ACIS-S3 0.5 - 5 keV image of A478
(background subtracted and exposure corrected) with isointensity contours
superposed. The source
was centered on S3 CCD. The contour levels increase by a factor of $\sqrt{2}$
from the outermost one (0.039 counts s$^{-1}$ arcmin$^{-2}$).
The image has been smoothed by gaussians with variable
$\sigma$, ranging from 3$''$ at the center to 20$''$ at the outskirts.
The small box at the center is the region shown on the right.
{\bf (b)}: the adaptively smoothed \chandra\ 0.5 - 5 keV image of the
central part of A478 (represented by the small box in the left image)
with 1.4 GHz radio contours superposed. The \chandra\ image was smoothed
to 3 - 4 $\sigma$ significance by CSMOOTH in CIAO. The color scale is linear
and ranges from 3.0 $\times$ 10$^{-4}$ to 2.5 $\times$ 10$^{-3}$ counts
s$^{-1}$ arcsec$^{-2}$. The radio contours
are in linear scale. The beam size of the radio observation is 1.5$''$.
The two X-ray cavities are clear and labeled.
    \label{fig:img:smo}}
\end{figure*}

\clearpage
\begin{figure*}
\vspace{-10cm}
  \centerline{\includegraphics[height=1.5\linewidth]{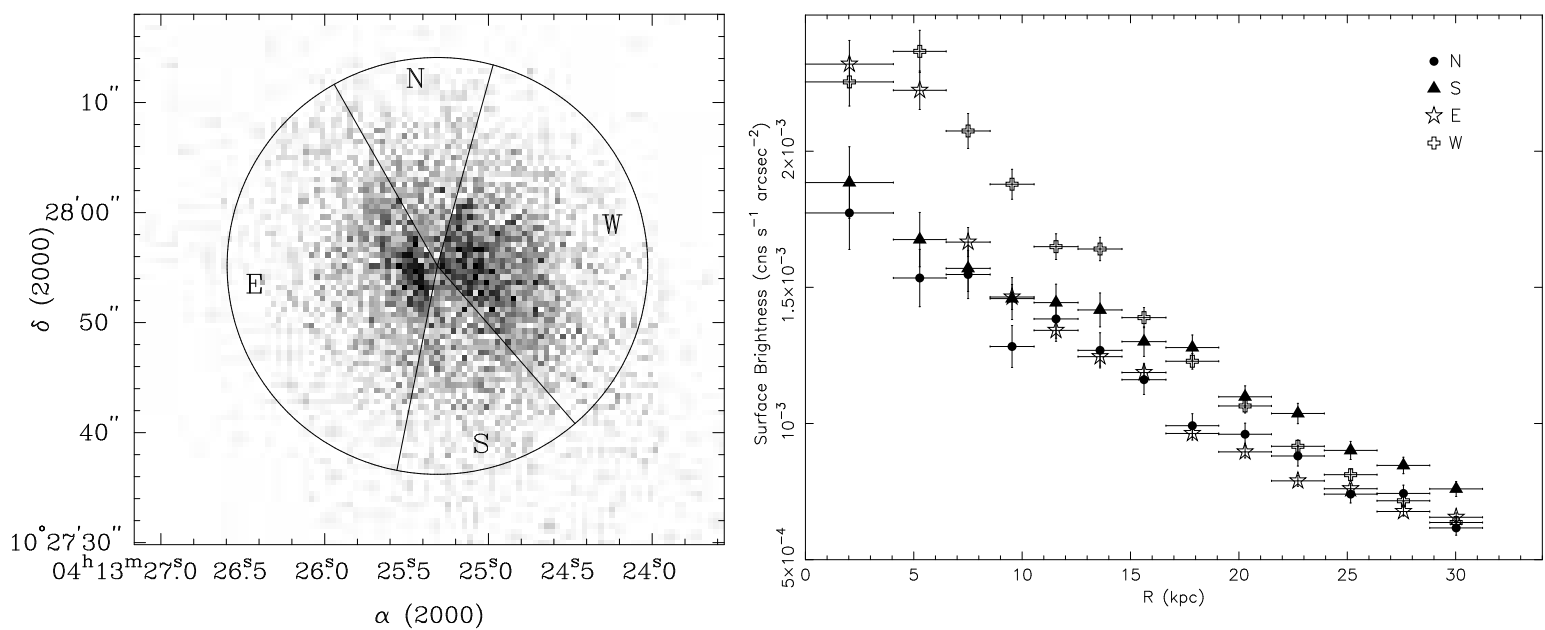}}
\vspace{-9cm}
  \caption{{\bf Left}: 0.5 - 5 keV photon image of the central part. The
four regions used to measure surface brightness profiles are shown;
{\bf Right}: Surface brightness profiles in four directions within the
inner 30 kpc. The depressions South and North of the center are significant.
The western peak is more extended than the eastern one.
    \label{fig:img:smo}}
\end{figure*}

\clearpage
\vspace{1cm}
\begin{inlinefigure}
  \centerline{\includegraphics[height=0.6\linewidth,angle=270]{fig3.ps}}
  \caption{The hardness ratio map of the central region of A478 (the same
region as Fig. 1b) superposed on the X-ray contours (in linear scale). The
hardness ratio is defined as (2.0 - 5.0 keV) / (0.5 - 2.0 keV). Thus, the lower
the value is, the cooler the temperature is. The typical statistical
uncertainty of the hardness ratio ranges from $\sim$ 0.02 at the center to
$\sim$ 0.03 near the edge of the field. Using the overall absorption value
for A478, a hardness ratio of 0.32 is expected for a gas with a 3.5 keV
temperature, while gas at 5.5 keV would produce a hardness ratio of 0.41.
    \label{fig:img:smo}}
\end{inlinefigure}

\clearpage

\vspace{-0.6cm}
\begin{inlinefigure}
  \centerline{\includegraphics[height=0.65\linewidth,angle=270]{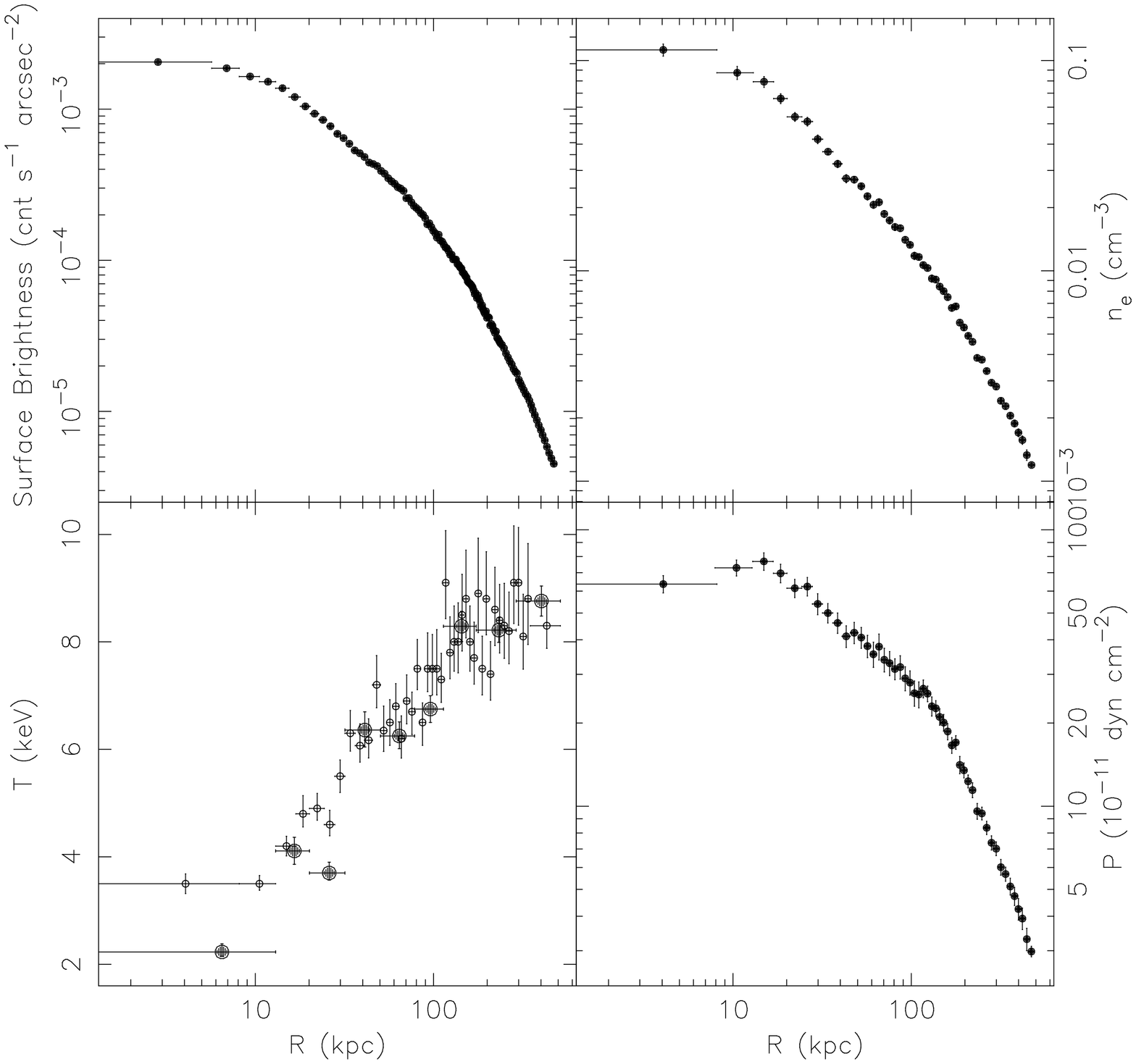}}
  \caption{Surface brightness (0.5 - 5 keV) (\emph{upper left}), electron
density (\emph{upper right}), temperature (\emph{lower left}), and pressure
(\emph{lower right}), as a function of radius. Only statistical uncertainties
are shown (1 $\sigma$ random errors).
In the temperature profile, the data points with small
circles are projected temperatures, while the data points with large filled
circles are deprojected temperatures. To derive the pressure profile,
we use the power-law fit of the temperature profile.
    \label{fig:img:smo}}
\end{inlinefigure}

\vspace{1.5cm}
\begin{inlinefigure}
  \centerline{\includegraphics[height=0.7\linewidth,angle=270]{fig5.ps}}
  \caption{{\bf Left}: the best-fit N$_{\rm H}$ with 90\% uncertainties in
9 radial bins to $\sim$ 500 kpc. The dashed line is the Galactic HI absorption
(1.53$\times$10$^{21}$ cm$^{-2}$). The observed
absorption is significantly higher than that value and peaks at
the center. {\bf Right}: the measured abundance with 90\% uncertainties
for each radial bin. The data points with stars are projected abundances,
while the data points with large filled circle are deprojected abundances.
The radial positions of the deprojected abundances are shifted a little bit
to allow the error bars to be better seen.
    \label{fig:img:smo}}
\end{inlinefigure}

\begin{inlinefigure}
  \centerline{\includegraphics[height=0.6\linewidth,angle=270]{fig6.ps}}
  \caption{X-ray cavity size vs. 1.4 GHz luminosity of the cD galaxy for
clusters with X-ray cavities detected. For each cluster, we plot the
larger cavity size. The cavity sizes are measured by eye and 10\% errors
are also shown. The 1.4
GHz luminosities for clusters were obtained from: A478 (our measurement);
A2052 (Zhao et al. 1993); A4059 (Heinz et al. 2002); A2597 (McNamara et al.
2001); Hydra A (Ekers \& Simkin 1983); Perseus (Fabian et al. 2000).
    \label{fig:img:smo}}
\end{inlinefigure}


\begin{references}

 \reference{} Allen, S. W., et al. 1993, MNRAS, 262, 901
 \reference{} Allen, S. W. 2000, MNRAS, 315, 269
 \reference{} Anders, E., \& Grevesse N. 1989, Geochimica et Cosmochimica Acta, 53, 197
 \reference{} Bahcall, N. A., \& Sargent, W. L. W. 1977, ApJ, 217, L19
 \reference{} Blanton, E., Sarazin, C., McNamara, B. R., \& Wise, M. W. 2001, ApJ, 558, L15
 \reference{} Boehringer, H., Voges, W., Fabian, A. C., Edge, A. C., Neumann, D. M. 1993, MNRAS, 264, 25
 \reference{} Churazov, E., et al. 2001, ApJ, 554, 261
 \reference{} David, L. P., et al. 2001, ApJ, 557, 546
 \reference{} Edge, A. C. 2001, MNRAS, 328, 762
 \reference{} Ekers, R. D., \& Simkin, S. M. 1983, ApJ, 265, 85
 \reference{} Fabian, A. C., et al. 2000, MNRAS, 318, L65
 \reference{} Fabian, A. C., et al. 2002, MNRAS, 331, 369
 \reference{} Heinz, S., Choi, Yun-Young, Reynolds, C. S., \& Begelman, M. C. 2002, ApJ, 569, L79
 \reference{} Hughes, J. P., \& Birkinshaw, M. 1998, ApJ, 501, 1
 \reference{} Johnstone, R. M., Fabian, A. C., Edge, A. C., \& Thomas, P. A. 1992, MNRAS, 255, 431
 \reference{} Markevitch, M., et al. 2000, ApJ, 541, 542
 \reference{} Markevitch, M., \& Vikhlinin, A. 2001, ApJ, 563, 95
 \reference{} Markevitch, M., et al. 2002, ApJ, in press (astro-ph/0209441)
 \reference{} Mason, B. S., Myers, S. T., \& Readhead, A. C. S. 2001, ApJ, 555, L11 (M01)
 \reference{} Mauskopf, P. D., et al. 2000, ApJ, 538, 505
 \reference{} McNamara, B. R., et al. 2000, ApJ, 534, L135
 \reference{} McNamara, B. R., et al. 2001, ApJ, 562, L149
 \reference{} Myers, S. T., et al. 1997, ApJ, 485, 1 (M97)
 \reference{} Plucinsky, P. P. et al. 2002, to appear in Astronomical Telescopes and Instrumentation 2002 (SPIE), (astro-ph/0209161)
 \reference{} Reese, E. D., et al. 2000, ApJ, 533, 38
 \reference{} Soker, N., Blanton, E., \& Sarazin, C. 2002, ApJ, 573, 533
 \reference{} White, D. A., et al. 1994, MNRAS, 269, 589
 \reference{} White, D. A. 2000, MNRAS, 312, 663
 \reference{} Zhao, J. H., Sumi, D. M., Burns, J. O., \& Duric, N. 1993, ApJ, 416, 51

\end{references}
\end{document}